\documentclass[12pt,letter]{article}
\pdfoutput=1
\usepackage{graphicx, epsfig, color,cite}
\usepackage{amsmath}
\usepackage{amssymb}
\usepackage{caption,subcaption,graphicx}
\usepackage[hidelinks]{hyperref}
\def\to{\rightarrow}

\def\bi{\begin{itemize}}
\def\ei{\end{itemize}}
\def\te{\tilde e}
\def\tU{\tilde U}

\def\tm{\tilde m}

\def\tu{\tilde u}

\def\tf{\tilde f}
\def\td{\tilde d}
\def\tQ{\tilde Q}

\def\tL{\tilde L}

\def\tst{\tilde t}

\def\tg{\tilde g}
\def\tnu{\tilde\nu}
\def\tell{\tilde\ell}
\def\tq{\tilde q}

\def\tw{\widetilde\chi^{\pm}}
\def\tz{\widetilde\chi^0}
\def\alt{\lesssim}
\def\agt{\gtrsim}
\def\be{\begin{equation}}  
\def\ee{\end{equation}}  
\def\bea{\begin{eqnarray}}  
\def\eea{\end{eqnarray}}

\begin{document}
\begin{titlepage}
\begin{flushright}
OU-HEP-190930
\end{flushright}

\vspace{0.5cm}
\begin{center}
{\Large \bf A landscape solution to the \\
SUSY flavor and CP problems\footnote{This paper is dedicated to
the memory of Ann Nelson, 
whose paper on Effective Supersymmetry, hep-ph/9607394, was an
inspiration for this work.}
}\\ 
\vspace{1.2cm} \renewcommand{\thefootnote}{\fnsymbol{footnote}}
{\large Howard Baer$^1$\footnote[1]{Email: baer@ou.edu },
Vernon Barger$^2$\footnote[2]{Email: barger@pheno.wisc.edu} and
Dibyashree Sengupta$^1$\footnote[3]{Email: Dibyashree.Sengupta-1@ou.edu}
}\\ 
\vspace{1.2cm} \renewcommand{\thefootnote}{\arabic{footnote}}
{\it 
$^1$Homer L. Dodge Department of Physics and Astronomy,
University of Oklahoma, Norman, OK 73019, USA \\[3pt]
}
{\it 
$^2$Department of Physics,
University of Wisconsin, Madison, WI 53706 USA \\[3pt]
}

\end{center}

\vspace{0.5cm}
\begin{abstract}
\noindent
In a fertile patch of the string landscape which includes the Minimal
Supersymmetric Standard Model (MSSM) as the low energy effective theory, 
rather general arguments from Douglas suggest a power-law statistical selection 
of soft breaking terms ($m_{soft}^n$ where $n=2n_F+n_D-1$ with $n_F$ 
the number of hidden sector $F$-SUSY breaking fields and $n_D$ the 
number of $D$-term SUSY breaking fields). 
The statistical draw towards large soft terms must be tempered by 
requiring an appropriate breakdown of electroweak (EW) symmetry with 
no contributions to the weak scale larger than a factor
2-5 of its measured value, lest one violates the (anthropic) atomic principle. 
Such a simple picture of stringy naturalness 
generates a light Higgs boson with mass $m_h\simeq 125$
GeV with sparticles (other than higgsinos) typically beyond LHC reach. 
Then we expect first and second generation matter scalars to be drawn 
independently to the tens of TeV regime where the upper cutoff arises 
from two-loop RGE terms which drive third generation soft masses towards
tachyonic values.
Since the upper bounds on $m_0(1,2)$ are the same for each generation, 
and flavor independent, 
then these will be drawn toward quasi-degenerate values.
This mechanism leads to a natural mixed decoupling/quasi-degeneracy 
solution to the SUSY flavor problem 
and a decoupling solution to the SUSY CP problem.
\end{abstract}
\end{titlepage}

\section{Introduction}
\label{sec:intro}

The emergence of the string landscape picture\cite{Bousso:2000xa,Susskind:2003kw,Douglas:2006es,Bousso:2004fc,Douglas:2019kus} provides so far the only plausible
mechanism for understanding the extreme suppression of the 
vacuum energy density of the universe $\rho_{vac}=\Lambda c^2/(8\pi G_N)\simeq (3\ meV)^4$ 
from its expected value $\sim m_P^4$ (over 120 orders of magnitude suppression).
Assuming a multiverse\cite{Linde:2015edk} with a huge 
(of order $10^{500}$\cite{Denef:2004ze} or far greater?\cite{Taylor:2015xtz}) 
assortment of vacua states with cosmological constant uniformly 
distributed across the decades, 
then those pocket universes with $\Lambda$ somewhat larger 
than our measured value would lead to such rapid expansion 
that galaxies wouldn't condense, and presumably observors wouldn't arise.
Weinberg used such reasoning to predict the value of $\Lambda$ to within 
a factor of several well before it was experimentally measured\cite{Weinberg:1987dv,Weinberg:1988cp}.

Given the success of the landscape in predicting $\Lambda$, can multiverse
arguments also be used to predict the scale of SUSY breaking\cite{Susskind:2004uv,Douglas:2004qg}?
A statistical approach to understanding the SUSY breaking scale has been 
advocated by Douglas\cite{Douglas:2004qg,Douglas:2012bu}. 
In this approach, naturalness is replaced by stringy naturalness\cite{Douglas:2004zg,Baer:2019cae} wherein observable ${\cal O}_2$ is more natural than 
observable ${\cal O}_1$ if more
{\it phenomenologically viable} vacua lead to ${\cal O}_2$ than to ${\cal O}_1$.
The key phrase  ``phenomenologically viable'' can be used here in an anthropic
sense, as in the case of the cosmological constant, in that such vacua 
lead to pocket universes that can admit life as we understand it.

Specifically, we might write the distribution of vacua as\cite{Douglas:2004qg}
\be
dN_{vac}[m_{hidden}^2,m_{weak},\Lambda ]
=f_{SUSY}(m_{hidden}^2 )\cdot f_{EWSB}\cdot f_{CC}\cdot dm_{hidden}^2
\label{eq:dNvac}
\ee
where $m_{hidden}$ is a mass scale associated with hidden sector SUSY breaking
which gives rise to (in gravity mediation, which is assumed here)\footnote{
In gauge mediation, typically the trilinear $A$ parameter $\sim 0$ so there
is little mixing in the stop sector, and consequently too light a
value for the SM-like Higgs boson $m_h$, unless soft terms have extremely 
large, unnatural values\cite{Arbey:2011ab,Draper:2011aa,Baer:2012uya,Baer:2014ica}. In gravity-mediation, since then we expect large
$A$ terms, there is no such problem to gain $m_h\simeq 125$ GeV
with natural soft term values under the $\Delta_{EW}$ finetuning measure\cite{Baer:2012up,Baer:2012cf}.} 
a gravitino mass $m_{3/2}\simeq m_{hidden}^2/m_P$ via the super-Higgs mechanism.
In such models, then we expect the appearance of soft SUSY breaking terms,
collectively denoted here as $m_{soft}$, of order $m_{soft}\sim m_{3/2}$\cite{Soni:1983rm,Kaplunovsky:1993rd,Brignole:1993dj}.

For the prior distribution $f_{SUSY}$, Douglas proposed on rather
general grounds a power law ansatz\cite{Douglas:2004qg,Susskind:2004uv} 
\be
f_{SUSY}(m_{hidden}^2)\sim (m_{hidden}^2)^{2n_F+n_D-1}
\ee
where $n_F$ is the number of hidden sector $F$-breaking fields and $n_D$
is the number of contributing $D$-breaking fields. 
This is reflective of general string theory models which typically contain of order 10 hidden sectors
some or all of which might contribute to SUSY breaking.
Only for $n_F=0$, $n_D=1$ would we obtain (the usually assumed) uniform
distribution of soft breaking terms. 
Already for $n_D=0$, $n_F=1$, we would expect a
{\it linear} statistical draw towards large soft terms. For more complicated
hidden sectors, then the statistical draw toward large soft terms 
would be even stronger. 

Early on, these considerations led to extensive
debate over whether to expect high scale or weak scale SUSY breaking\cite{Douglas:2004qg,Susskind:2004uv,Dine:2004is}.
Such debate was in part predicated on the influence of 
cosmological constant selection on the SUSY breaking scale. 
Initial expectations were that $f_{CC}\sim\Lambda /m_{hidden}^4$.
Following Douglas\cite{Douglas:2004qg}, the consensus emerged that 
$f_{CC}$ would be independent of
the SUSY breaking sector, and that $f_{CC}\sim \Lambda /m_{string}^4$.

The third element in Eq. \ref{eq:dNvac} is $f_{EWSB}$. 
This function contains any anthropic requirements. 
For the case of SUSY, it also depends on the anticipated solution to the 
SUSY $\mu$ problem: why is the SUSY conserving $\mu$ parameter of order the weak
scale rather than the Planck scale\cite{Bae:2019dgg}? 
Here, we will assume a {\it natural} solution to the SUSY $\mu$ problem, 
{\it i.e.} that $|\mu |\sim m_{weak}$. If $|\mu|\gg m_{weak}$, then 
some finetuning would be required to gain a value of $m_{weak}$ close to
the 100 GeV scale. Such finetuning requires a tiny range of compensating
opposite-sign soft terms to maintain the weak scale not-too-far from its measured value\cite{Baer:2019cae}. And as shown by nuclear physics calculations of 
Agrawal {\it et al.}\cite{Agrawal:1998xa}, a pocket universe value of $m_{weak}$ 
displaced by a factor 2-5 from our measured value would lead to
catastrophes in nuclear physics that would violate the atomic principle.

The magnitude of the weak scale is related to SUSY Lagrangian parameters
via the scalar potential minimization condition
\be
m_Z^2/2=\frac{m_{H_d}^2+\Sigma_d^d-(m_{H_u}^2+\Sigma_u^u)\tan^2\beta}
{\tan^2\beta -1}-\mu^2\simeq -m_{H_u}^2-\Sigma_u^u-\mu^2
\label{eq:mzs}
\ee
where the $\Sigma_u^u$ and $\Sigma_d^d$ terms contain a large sum of radiative corrections (for expressions, see the Appendix to \cite{Baer:2012cf}). 
In fact, the electroweak finetuning measure $\Delta_{EW}$\cite{Baer:2012up,Baer:2012cf} 
conservatively requires that the {\it weak scale} terms on the right-hand-side
of Eq. \ref{eq:mzs} be comparable to the observed value $m_Z^2/2$ on the 
left-hand-side. This is a manifestation of the notion of 
{\it practical naturalness}\cite{Baer:2019xww}:
that various additive contributions to any observable should be comparable to 
or less than that observable: if not, then (implausible) finetunings are 
required to enforce the observable at its measured value. 
In most SUSY phenomenology papers, the measured value of $m_Z$ 
is used to fix (finetune) the value of $\mu^2$. 
In our approach, since $\mu$ is already fixed at a natural
value due to the solution to the SUSY $\mu$ problem, 
then $m_Z$ is left variable. 
We denote the various pocket universe values of the $Z$-mass as 
$m_Z^{PU}$ which is different from the measured value in our universe\cite{Baer:2016lpj,Baer:2017uvn}.
With $\mu$ fixed, then different statistical entries for the soft terms 
will determine an associated pocket universe value for the weak scale, 
and consequently for $m_Z^{PU}$.

An initial guess for this term\cite{ArkaniHamed:2004fb} was $f_{EWSB}\sim (m_{weak}/m_{soft})^2$ which follows the gross behavior
of finetuning measures $\Delta_{BG}$\cite{bg} ($\Delta_{EW}$) which compare the largest
high scale (weak scale) SUSY breaking contribution to the size of the
weak scale itself: then the ansatz for $f_{EWSB}$ rewards vacua with soft terms
that are closest to the magnitude of the weak scale itself.

As pointed out in Ref. \cite{Baer:2016lpj,Baer:2017uvn}, this ansatz
fails in five cases (and a sixth case will be discussed later in 
Sec. \ref{sec:heavy}).
\begin{enumerate}
\item Very large trilinear soft terms lead to charge-or-color breaking 
(CCB) vacua. 
Such vacua are unlikely to support the existence of 
atoms\footnote{This is known as the {\it atomic principle}: life as we know
it seems to require the existence of atoms and molecules as exist in 
{\it e.g.} our universe.} (the atomic principle) and hence life as we know it.
Such CCB minima must be vetoed and not merely penalized by a statistical factor.
\item If other soft terms such as $m_{H_u}^2$ are too large, then they are 
not driven negative at $Q=m_{weak}$ and EW symmetry is not even broken.
Such vacua must also be vetoed.
\item For small (seemingly more natural) values of $m_{H_u}^2$, then
$m_{H_u}^2$ is driven to large negative values, resulting in too large 
of values of $m_Z^{PU}$, in violation of Agrawal {\it et al.} limits.
As $m_{H_u}^2(\Lambda )$ increases, its weak scale value {\it decreases}
(radiatively-driven naturalness) resulting in a more natural theory with
$m_Z^{PU}$ close to the measured value in our universe.
\item As the trilinear soft term $A_0$ increases (seemingly more unnatural), 
then large cancellations in $\Sigma_u^u (\tst_{1,2})$ render these contributions more natural, and $m_Z^{PU}$ closer to our measured value.
\item Even in the event of appropriate EW symmetry breaking, the
ansatz $f_{EWSB}\sim (m_{weak}/m_{soft})^2$ penalizes but does not forbid vacua 
with too large a value of $m_Z^{PU}$. 
Surviving vacua with $m_Z^{PU}\agt (2-5)m_Z^{meas}$ must be vetoed since these 
would contradict the nuclear physics analyses of Agrawal {\it et al.}\cite{Agrawal:1998xa}. 
\end{enumerate}
To ameliorate this situation, it was proposed in Ref's \cite{Baer:2016lpj,Baer:2017uvn} 
to instead veto any non-standard EW vacua and also to veto any vacua with too large
a value of $m_Z^{PU}$ greater than a factor four larger than our measured value.
For a fixed natural value of $\mu$, this latter condition corresponds 
to vetoing pocket universes with $\Delta_{EW}>30$. Thus, we also implement
\be
f_{EWSB}=\Theta (30-\Delta_{EW} ) .
\ee

By scanning over models such as NUHM2\cite{nuhm2} or NUHM3 which allow for
an input $\mu$ parameter, with soft terms generated according to
$m_{soft}^n$ for $n=1$ and 2, along with the anthropic vetos from $f_{EWSB}$, 
then the following features were found\cite{Baer:2017uvn}:
\begin{itemize}
\item A statistical peak was found at $m_h\simeq 125\pm 2$ GeV. 
This is easy to understand: we are selecting for soft terms as large as possible subject to appropriate EWSB and a value of $m_Z^{PU}\alt 4 m_Z^{meas}$. 
This also selects for large (but not so large as to lead to CCB minima) 
$A_0$ terms which increase top squark mixing and lift $m_h$ up to
the vicinity of 125 GeV.
\item The probability distribution $dP/dm_{\tg}$ yields a value
$m_{\tg}\sim 4\pm 2$ TeV, safely above LHC2 limits.
\item The light top squark is lifted to $m_{\tst_1}\sim 1.5\pm 0.5$ TeV, 
also safely above LHC Run 2 limits.
\item Light higgsinos $\tw_{1}$ and $\tz_{1,2}$ with mass $\sim \mu\sim 200\pm 100$ GeV. The mass gap is $m_{\tz_2}-m_{\tz_1}\sim 7\pm 3$ GeV. 
Thus, higgsino pair production signals should ultimately show up at LHC14
 via $pp\to\tz_1\tz_2$ production followed
by $\tz_2\to\ell^+\ell^-\tz_1$ decay with $m(\ell^+\ell^- )<(7\pm 3)$ GeV 
once sufficient luminosity is gained\cite{SDLJMET,Sirunyan:2018iwl,atlas:SDLJMET}.
\item First and second generation matter scalars (squarks and sleptons) are
pulled up to $m(\tq ,\tell)\sim 20\pm 10$ TeV.
\end{itemize}

The present paper focuses on this latter point. 
Apparently, with first and second generation matter scalars being pulled up 
to the multi-TeV regime, then one is also being pulled up to a 
potential {\it decoupling} solution to the SUSY flavor and CP problems.
The question is: how does this decoupling arise, and is it enough to 
actually solve these two SUSY issues?

\section{Living dangerously with heavy sfermions}
\label{sec:heavy}

In Sec. \ref{sec:intro}, we emphasized that Douglas' general 
stringy considerations imply a statistical draw towards large soft terms. 
However, the soft terms cannot become arbitrarily large without leading 
to non-standard EW vacua or else too large of a value of pocket universe 
weak scale $m_Z^{PU}$: such vacua must be anthropically vetoed. 

Here, we concern ourselves with the upper bound on matter sfermion masses 
for the first two generations, which we label according to high-scale soft term
values $m_0(1)$ and $m_0(2)$. 
For simplicity, we will assume all high scale matter sfermion masses 
within a single generation are degenerate (as is expected 
in models containing some remnant $SO(10)$ GUT symmetry). 
These could be placed for context within the
$i$-extra parameter non-universal Higgs models\cite{nuhm2} (NUHMi, $i=2-4$).
In NUHM2 $m_0(1)$ = $m_0(2)$ = $m_0(3)$ while 
in NUHM3 $m_0(1)$ = $m_0(2)$ $\neq$ $m_0(3)$.  
Here, NUHM4 is considered since we are allowing for 
splittings between first and second generation masses (as well as the third)
{\it i.e.}  $m_0(1)$ $\neq$ $m_0(2)$ $\neq$ $m_0(3)$. 
But we will also allow for the presence of off-diagonal
soft term masses. To make contact with general constraints from SUSY flavor 
and CP violating processes, as presented for instance in 
Ref's \cite{Gabbiani:1988rb}, \cite{Hagelin:1992tc}, \cite{Gabbiani:1996hi} and \cite{Misiak:1997ei}. 
we will work within the superCKM mass basis
wherein the quark and lepton mass matrices are diagonal but the squark 
and slepton mass matrices are not yet diagonalized.

From a scan over NUHM3 parameter space in Ref. \cite{Baer:2017uvn}, 
it was found that the statistical distribution of first/second 
generation sfermion masses for $n=1$ or 2 was peaked around 
$m_{\tf}\sim 20$ TeV but with tails extending as far as 40 TeV.
What sets the upper bound for such sfermion masses? 

At first sight, the $\Sigma_u^u$ and $\Sigma_d^d$ terms contain 
first/second generation $D$-term contributions to the EW scale.
For first/second generation sfermions, neglecting the small Yukawa
couplings, we find the contributions 
\bea \Sigma_{u,d}^{u,d}(\tf_{L,R}) = \mp
\frac{c_{col}}{16\pi^2}F(m_{\tf_{L,R}}^2)\left(-4g_Z^2(T_3-Q_{em}x_W) \right),
\label{eq:sigud}
\eea 
where $T_3$ is the weak isospin, $Q_{em}$ is the electric charge
assignment (taking care to flip the sign of $Q_{em}$ for right-sfermions),
$c_{col}=1 (3)$ for color singlet (triplet) states, 
$x_W\equiv\sin^2\theta_W$ and where 
\be
F(m^2)= m^2\left(\log\frac{m^2}{Q^2}-1\right) .  
\ee 
We adopt an optimized scale choice 
$Q^2 = m_{\rm SUSY}^2 \equiv
m_{\tst_1}m_{\tst_2}$.\footnote{The optimized scale choice is chosen to
minimize the log contributions to $\Sigma_u^u(\tst_{1,2})$ which occur
to all orders in perturbation theory.}  The explicit first generation
squark contributions to $\Sigma_u^u$ (neglecting the tiny Yukawa
couplings) are given by
\bea
\Sigma_u^u (\tu_L) &=& \frac{3}{16\pi^2}F(m_{\tu_L}^2)\left(-4g_Z^2
(\frac{1}{2}-\frac{2}{3} x_W)\right)\nonumber \\
\Sigma_u^u (\tu_R) &=& \frac{3}{16\pi^2}F(m_{\tu_R}^2)\left(-4g_Z^2 (\frac{2}{3}x_W)\right)\\
\Sigma_u^u (\td_L) &=& \frac{3}{16\pi^2}F(m_{\td_L}^2)\left(-4g_Z^2
(-\frac{1}{2}+\frac{1}{3}x_W)\right)\nonumber \\
\Sigma_u^u (\td_R) &=& \frac{3}{16\pi^2}F(m_{\td_R}^2)\left(-4g_Z^2
(-\frac{1}{3}x_W)\right) . \nonumber
\label{eq:Ssquarks}
\eea
These contributions, arising from electroweak $D$-term contributions to
masses, are frequently neglected since the various
contributions cancel amongst themselves in the limit of mass degeneracy
due to the fact that weak isospins and electric charges (or weak
hypercharges) sum to zero in each generation.  However, if squark and
slepton masses are in the multi-TeV regime but are {\it non-degenerate}
within each generation, then the contributions may be large and
non-cancelling.  In this case, they may render a theory which is
otherwise considered to be natural, in fact, unnatural.

The first generation slepton contributions to $\Sigma_u^u$ are given by
\bea
\Sigma_u^u (\te_L) &=& \frac{1}{16\pi^2}F(m_{\te_L}^2)\left(-4g_Z^2
(-\frac{1}{2}+x_W)\right) \nonumber \\
\Sigma_u^u (\te_R) &=& \frac{1}{16\pi^2}F(m_{\te_R}^2)\left(-4g_Z^2 (-x_W)\right)\\
\Sigma_u^u (\tnu_L) &=& \frac{1}{16\pi^2}F(m_{\tnu_{eL}}^2)\left(-4g_Z^2
(\frac{1}{2})\right) ; \nonumber
\label{eq:Ssleptons}
\eea
these may also be large for large $m_{\tell}^2$ although again they 
cancel amongst themselves in the limit of slepton mass degeneracy.

In our evaluation of $\Delta_{EW}$, in fact we sum all contributions 
from a complete generation before including them into $\Delta_{EW}$. 
This allows for complete $D$-term cancellations in the limits of weak scale
sfermion degeneracy. Of course, the sfermions are not completely
degenerate at the weak scale even if they begin as degenerate at the high scale
$Q\equiv m_{GUT}$ due at least to weak scale $D$-term contributions to 
their masses. We have evaluated these contributions and find they lead
to upper bounds on $m_0(1,2)\alt 5000$ TeV for $\Delta_{EW}<30$, so that these
$D$-terms do not set the upper limits on first/second generation sfermion 
masses.

A stricter constraint on first/second generation sfermion masses from 
the landscape comes from 2-loop RGE contributions to the running of sfermion 
masses.
The form of the two loop RGEs for sfermion masses is given by
\be
\frac{dm_i^2}{dt}=\frac{1}{16\pi^2}\beta_{m_i^2}^{(1)}+
\frac{1}{(16\pi^2)^2}\beta_{m_i^2}^{(2)},
\label{eq:RGE}
\ee
where $t=\ln Q$,
$i=Q_j,\ U_j,\ D_j,\ L_j$ and $E_j$, and $j=1-3$ is a generation index.
The one loop $\beta$-function for the evolution of {\it third} generation
scalar masses depends only on third generation and Higgs scalar masses 
and on the gaugino masses.
The two loop terms are formally suppressed relative to one loop terms by
the square of a coupling constant as well as an additional loop
factor of $16\pi^2$. However, these two loop terms include contributions
from {\it all} scalars. 
Specifically, the two loop $\beta$ functions include\cite{Martin:1993zk}
\be
\beta_{m_i^2}^{(2)}\ni a_ig_3^2\sigma_3+b_ig_2^2\sigma_2+c_ig_1^2\sigma_1,
\ee
where 
\bea
\sigma_1 &=& {1\over 5}g_1^2\{3(m_{H_u}^2+m_{H_d}^2)+Tr[{\bf m}_Q^2+
3{\bf m}_L^2+8{\bf m}_U^2+2{\bf m}_D^2+6{\bf m}_E^2]\},\nonumber \\
\sigma_2 &=& g_2^2\{m_{H_u}^2+m_{H_d}^2+Tr[3{\bf m}_Q^2+
{\bf m}_L^2]\},\ \ \ \ {\rm and} \nonumber\\
\sigma_3 &=& g_3^2Tr[2{\bf m}_Q^2+{\bf m}_U^2+{\bf m}_D^2],\nonumber
\eea
and the ${\bf m}_i^2$ are squared mass matrices in generation space.
The numerical coefficients $a_i$, $b_i$ and $c_i$ are related to the quantum
numbers of the scalar fields, but are all positive quantities. 

Thus, incorporation of multi-TeV masses for the first and second generation
scalars leads to an overall positive, {\it possibly dominant},
contribution to the slope of third generation soft mass trajectories 
versus energy scale.
Although formally a two loop effect, 
the smallness of the couplings is compensated by the 
much larger values of masses of the first two generations of scalars.
In running from $m_{\rm GUT}$ to $m_{\rm weak}$,
this results in an overall {\it reduction} in third generation scalar masses.
In fact, this effect was argued in Ref.~\cite{ArkaniHamed:1997ab} to lead to 
violation of naturalness constraints from a decoupling solution 
to the SUSY flavor problem.
It was also used in Ref's \cite{Baer:2000xa} and \cite{Baer:2001vw}
to generate SUSY models with an inverted scalar mass hierarchy
to reconcile naturalness with a decoupling solution to the
SUSY flavor and CP problems along the lines of 
``effective supersymmetry''\cite{Cohen:1996vb}.
For values of sfermion masses which fall short of tachyonic, a sort of
see-saw effect amongst scalar masses occurs: the higher the value of
first and second generation scalar masses, the larger will be the 
two loop suppression of third generation and Higgs scalar masses.
In this class of models, first and second generation scalars
with masses of order $10-40$~TeV may co-exist with TeV-scale
third generation scalars, thus giving a very large suppression to
both FCNC and $CP$ violating processes while driving third generation
sfermions to natural values.

In the context of our string landscape picture, this is yet another example of
living dangerously\footnote{Arkani-Hamed and Dimopoulos\cite{ArkaniHamed:2005yv}  state:
``anthropic reasoning leads to the conclusion that we live dangerously close to
violating an important but fragile feature of the low-energy world...'', in this
case, appropriate electroweak symmetry breaking.}, 
wherein soft terms are pulled to 
large values which actually {\it increases the naturalness of the 
theory} so long as we stop short of impending disaster: which 
in this case would be that huge first/second generation sfermion 
masses might drive third generation masses tachyonic leading to CCB vacua.

\begin{figure}[!htbp]
\begin{center}
\includegraphics[height=0.42\textheight]{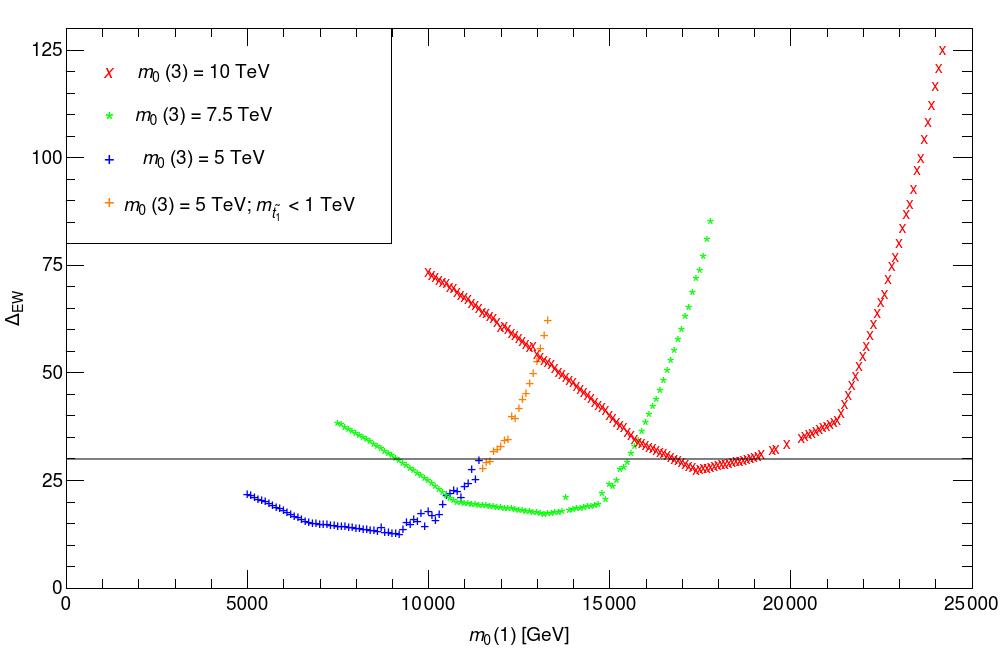}
\caption{We plot the value of $\Delta_{EW}$ vs. $m_0(1,2)$ 
for $m_0(3)=5,$ 7.5 and 10 TeV and $m_{1/2}=1200$ GeV, $A_0=-1.6 m_0(3)$ 
and $\tan\beta =10$ with $\mu=200$ GeV and $m_A=2000$ GeV.
\label{fig:m012}}
\end{center}
\end{figure}

The situation is illustrated in Fig. \ref{fig:m012} where we adopt the 
NUHM3 model to plot the value of $\Delta_{EW}$ versus $m_0(1,2)$ for
$m_{1/2}=1200$ GeV, $A_0=-1.6 m_0(3)$ 
and $\tan\beta =10$ with $\mu=200$ GeV and $m_A=2000$ GeV.
We also take $m_0(3)=5$, 7.5 and 10 TeV 
(blue/orange, green and red curves, respectively).
From the plot we see that as $m_0(1,2)$ increases, the models are driven 
to greater naturalness in that third generation soft terms are driven 
to smaller values by large two-loop RGE contributions.
As $m_0(1,2)$ increases even further, then cancellations with the 
$\Sigma_u^u(\tst_{1,2})$ terms are disrupted and the models again become
more unnatural, leading to too large of contributions to the 
pocket universe weak scale $m_Z^{PU}$. For even higher $m_0(1,2)$ values, 
then the top squark soft term $m_{\tst_R}^2$ is driven tachyonic 
leading to CCB vacua.

An important point is that for particular parameter values, 
we do gain an upper bound on first/second generation soft terms. 
The upper bound changes within parameter space variation, but
depends only on gauge quantum numbers, so it is the same for both 
generations one and two. Thus, the first and second generation
soft masses are pulled to large values by the landscape, 
but with the same upper bounds. This means that for strong enough pull, 
then $m_0(1)$ and $m_0(2)$ will be pulled to similar upper limits. If the
pull is strong enough, they will be pulled towards quasi-degeneracy, which
helps, along with decoupling,  to solve the SUSY flavor problem.

\section{SUSY flavor and CP problems}
\label{sec:flavorCP}

\subsection{Flavor}
\label{ssec:flavor}

In the SM, a fourth quark, charm, was posited in order to suppress
flavor changing neutral current (FCNC) processes, for which there were strict 
limits\cite{gim}. In a successful application of practical naturalness, 
Gaillard and Lee\cite{gl} required the charm-quark box diagram contribution 
to the $m_{K_L}-m_{K_S}\equiv \Delta m_K$ mass difference to be less than the 
measured value of $\Delta m_K$ itself: this lead to the successful 
prediction that $1\ {\rm GeV}<m_c<2$ GeV shortly before the charm quark 
discovery.

By supersymmetrizing the SM into the MSSM, then many new parameters
are introduced, mainly in the soft SUSY breaking sector\cite{ds}.
These include sfermion mass matrices 
\be
{\cal L}_{soft}\ni -\tf_i^\dagger ({\bf m_{f}^2})_{ij}\tf_j
\label{eq:m}
\ee
where $i$ and $j$ are generation indices $i,j=1-3$ and the sfermion
index $\tf$ runs over the various matter superfields $\hat{Q}$, 
$\hat{U}^c$, $\hat{D}^c$, $\hat{L}^c$ and $\hat{E}^c$ in the notation of
Ref. \cite{wss}. There are also trilinear soft terms that can contribute 
to flavor violation:
\be
{\cal L}_{soft}\ni ({\bf a}_u)_{ij}\epsilon_{ab}\tQ_i^aH_u^b\tu_{Rj}^\dagger 
+({\bf a}_d)_{ij}\tQ_i^a H_{da}\td_{Rj}^\dagger +
+({\bf a}_e)_{ij}\tL_i^a H_{da}\te_{Rj}^\dagger +h.c.
\ee
In gravity mediation, the trilinears are expected to be proportional to 
the corresponding Yukawa couplings so that these terms are small for 
first/second generation values. 
We will thus focus mainly on the mass matrices in Eq.~\ref{eq:m}.

In the superCKM basis, the $6\times 6$ sfermion mass matrices 
are built out of $3\times 3$ $LL$, $RR$, $LR$ and $RL$ 
sub-matrices which have the form 
{\it e.g.}\footnote{For a more detailed review, see Ref. \cite{Misiak:1997ei}.}
\be
({\bf m}_{\tf}^2)_{LL}= \left(\begin{array}{ccc} (m_{f 1}^2)_{LL} & 
(\Delta^{f}_{12})_{LL} & (\Delta^{f }_{13})_{LL} \\
(\Delta^{f}_{21})_{LL} & (m_{f 2}^2)_{LL} & (\Delta^{f }_{23})_{LL} \\
(\Delta^{f}_{31})_{LL} & (\Delta^{f}_{32})_{LL} & (m_{f 3}^2)_{LL}
\end{array}\right)
\ee
with $(m_{\tU}^2)_{LL}=V_L^u{\bf m}_Q^2V_L^{u\dagger}$, 
$(m_{\tU}^2)_{RR}=V_R^u{\bf m}_U^{2T}V_R^{u\dagger}$ and 
$(m_{\tU}^2)_{LR}=-\frac{v\sin\beta}{\sqrt{2}}V_L^u{\bf a}_U^*V_R^{u\dagger}$
{\it etc.} and where the CKM matrix is given by $V_{KM}=V_L^uV_L^{d\dagger}$. 
For mass matrices proportional to the unit matrix 
${\bf m}_{\tf}^2= m_{\tf}^2{\bf 1}$ 
(flavor universality), then no flavor-changing transitions are allowed
and the SUSY flavor problem is solved. But for gravity-mediation, no known
principles enforce flavor universality because the transformation that 
diagonalizes the quark mass matrices does not simultaneously 
diagonalize the corresponding squark mass squared matrices. 
In that case, then the off-diagonal mass matrix contributions $\Delta_{ij}^f$
may contribute to FCNC processes via mass insertions, and furthermore, 
non-degenerate diagonal terms can also lead to FCNC 
effects\cite{Ellis:1981ts}.
Constraints on the off-diagonal terms are typically listed in terms
of dimensionless quantities 
$(\delta^f_{ij})_{LL,RR,LR,RL}\equiv\frac{(\Delta^f_{ij})_{LL,RR,LR,RL}}{\tm^2}$
where the $\tm$ represent an averaged sfermion mass for the
corresponding mass matrix.

\begin{figure}[!htbp]
\begin{center}
\includegraphics[height=0.42\textheight]{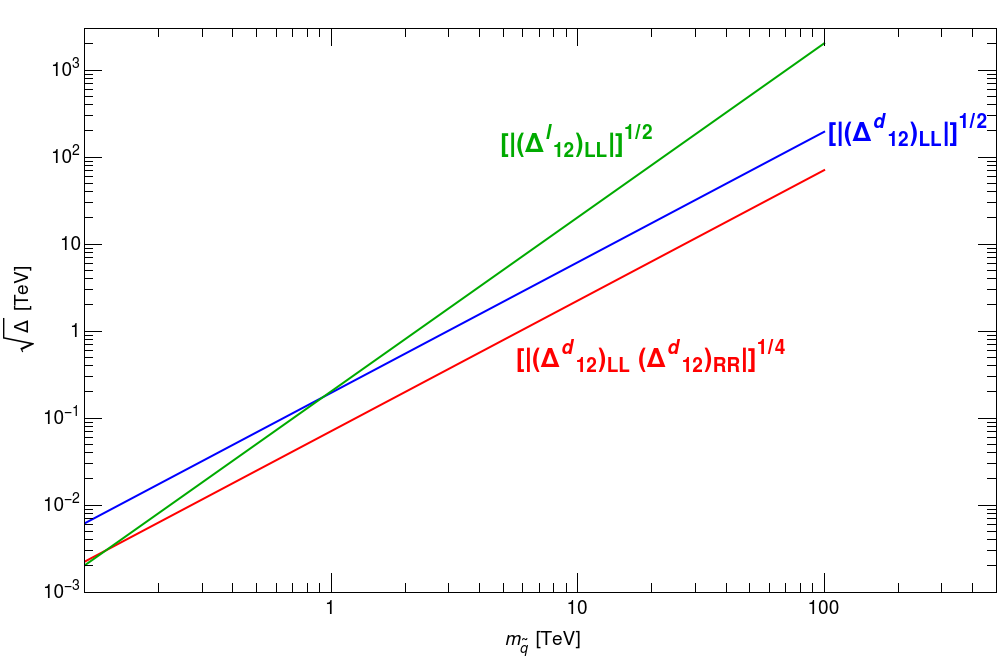}
\caption{Upper limits on off-diagonal squark mass terms from
$\Delta m_K$ constraints (blue and red) and off-diagonal slepton masses from
$BF(\mu\to e\gamma )$ (green).
\label{fig:Delta}}
\end{center}
\end{figure}

First we concentrate on limits for flavor-changing off-diagional 
mass matrix elements as they vary from the weak scale on into the
decoupling regime. 
In Fig. \ref{fig:Delta}, we list the most restrictive limits on several
$\Delta_{ij}$ quantities arising from $\Delta m_K$ constraint\cite{Donoghue:1983mx,Duncan:1983wz,Bouquet:1984pp} and also
from updated branching fraction limits on $\mu\to e\gamma$ decay:
$BF(\mu\to e\gamma )<4.2\times 10^{-13}$ at 90\% CL\cite{meg}.
We plot Fig. \ref{fig:Delta} for $m_{\tg}^2\sim .3m_{\tq}^2$ for
$\Delta m_K$ constraints and $m_{\tz_1}^2=0.3 m_{\tell}^2$ although the
constraints only depend weakly on these mass 
ratios\cite{Gabbiani:1996hi,Hagelin:1992tc}.
From Fig. \ref{fig:Delta}, we see that for sfermion masses of order the
weak scale $\sim 100$ GeV, then the updated $\mu\to e\gamma$ 
branching fraction now slightly pre-empts the $\Delta m_K$ 
constraints although all require off-diagonal mass terms less than
$1-10$ GeV. These limits exemplify the SUSY flavor problem from days 
gone by when sparticles were expected to occur around the weak scale.
As $m_{\tf}$ increases, then the restrictions on off-diagonal 
masses become increasingly mild, thus illustrating the onset of the 
decoupling solution to the SUSY flavor problem. 
For large sfermion masses, then the $\Delta m_K$ constraint is again most 
confining. For $m_{\tf}\sim 10$ TeV, the off-diagonal masses
are constrained to be $\alt 1-10$ TeV while for landscape SUSY masses, 
where first/second geenration sfermions are expected in the $20-30$ TeV range, 
then the off-diagonal limits are $\alt 5-50$ TeV. 
Such values are only mildly suppressed compared to the average 
squark/slepton masses although one must proceed into the 
$m_{\tf}\sim 100$ TeV range for unfettered flavor violation\cite{ArkaniHamed:1997ab}.

Along with limits on off-diagonal mass matrix terms, to achieve flavor
universality one needs degeneracy on the diagonal. 
Limits on degeneracy have been computed in 
Misiak {\it et al.}\cite{Misiak:1997ei}. 
From the $\Delta m_K$ constraint, for the first two generations of squarks 
these amount to
\be
|m_{\tq 1}-m_{\tq 2}|\alt 2 m_c m_{\tq}^2/ m_W^2 
\ee
for both up and down squarks. Thus, for sparticle masses of order $m_W$, 
splittings of only a few GeV are allowed and we must be in a state of 
near degeneracy. As $m_{\tq}$ increases, then these bounds become much weaker.

\begin{figure}[!htbp]
\begin{center}
\includegraphics[height=0.24\textheight]{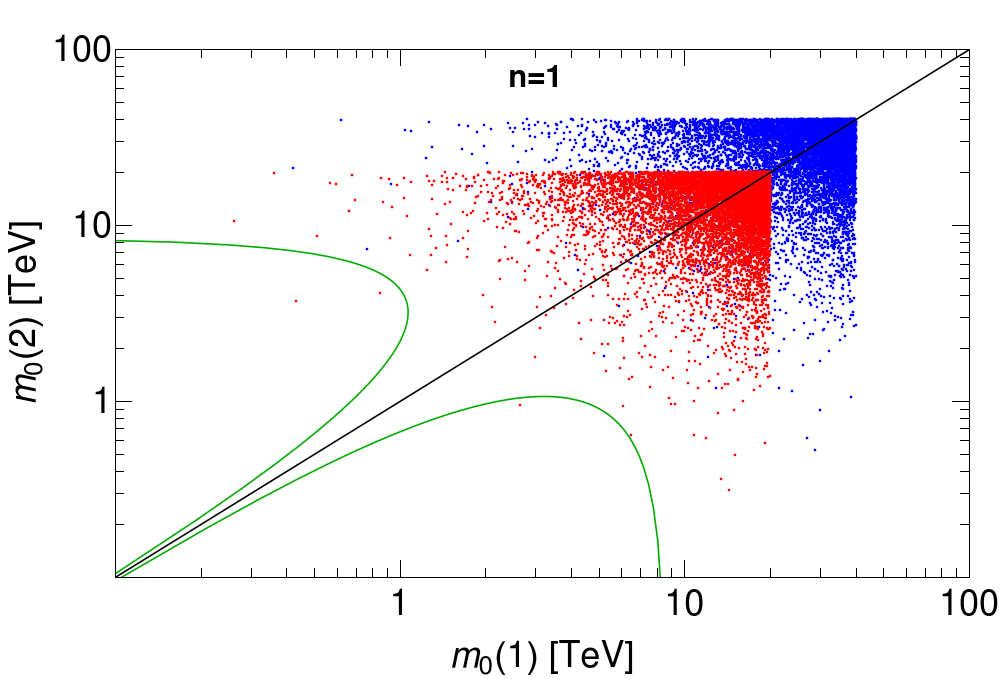}
\includegraphics[height=0.24\textheight]{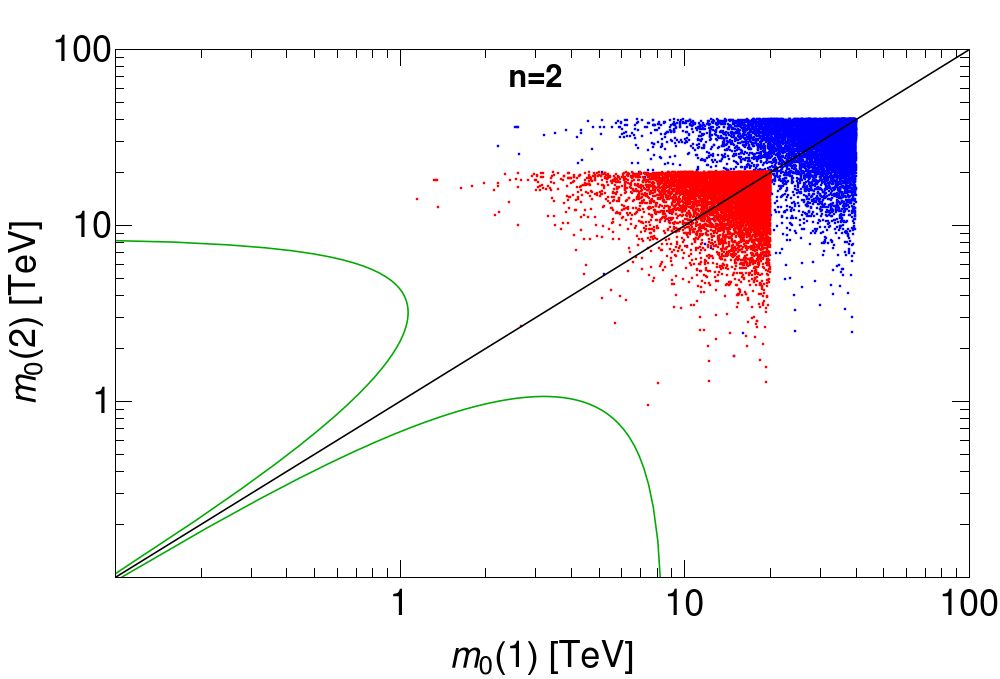}\\
\includegraphics[height=0.24\textheight]{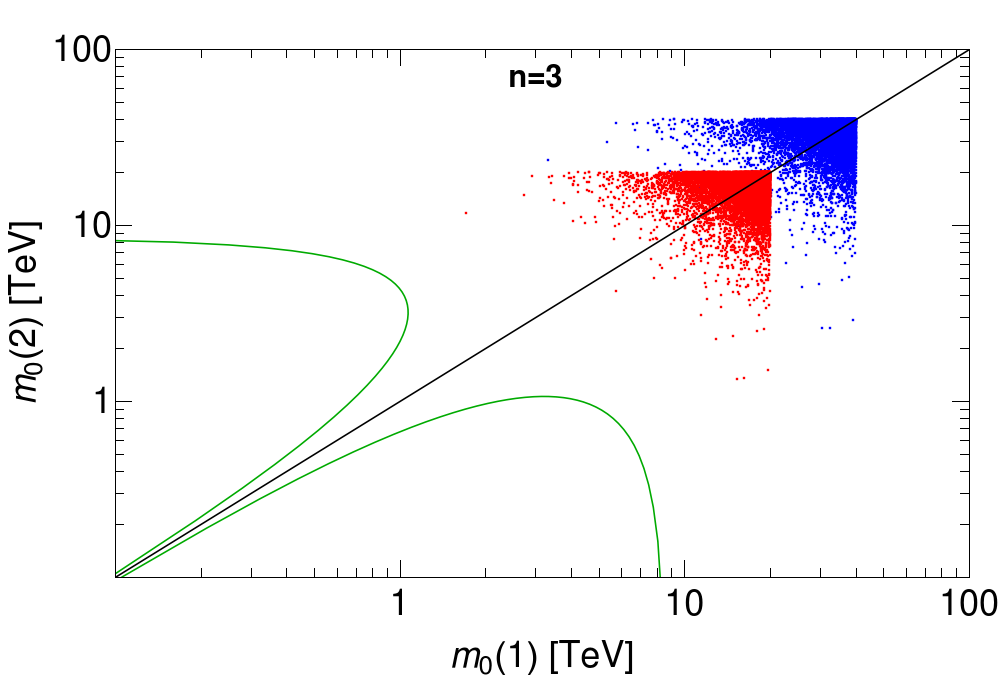}
\includegraphics[height=0.24\textheight]{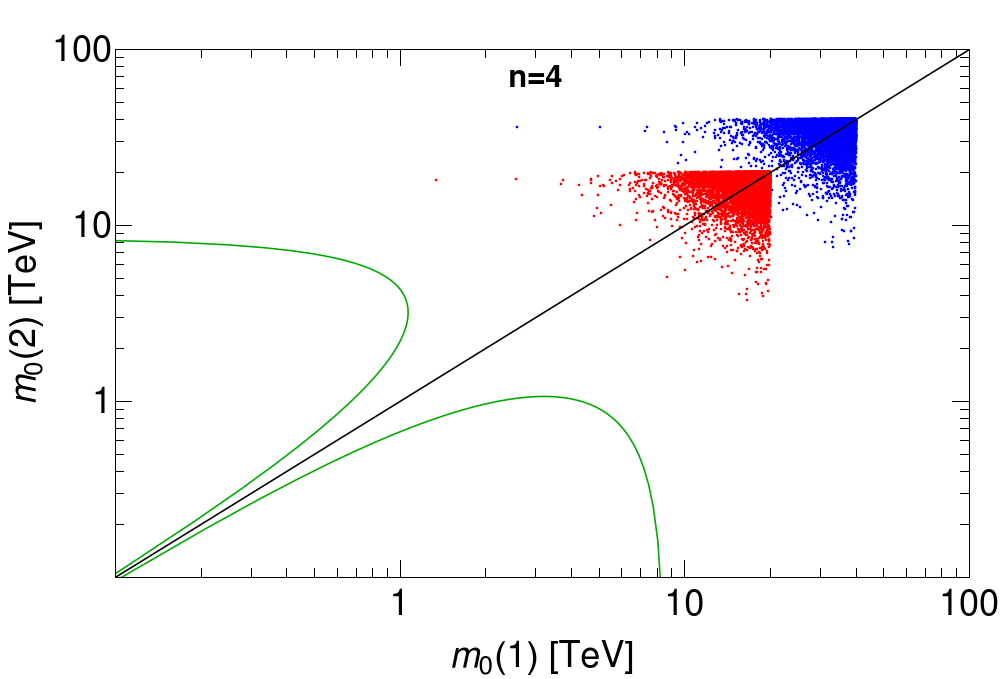}
\caption{The values of $m_0(2)$ vs. $m_0(1)$ from an {\it a}) $n=1$,
 {\it b}) $n=2$, {\it c}) $n=3$ and {\it d}) $n=4$,
statistical selection of first and second generation matter scalar soft terms.
The lower-left of green curves is excluded while red points denote soft terms scanned up to 20 TeV while blue points show points scanned up to 40 TeV.
\label{fig:m01vsm02}}
\end{center}
\end{figure}

The situation is shown in Fig. \ref{fig:m01vsm02} where we plot the
GUT scale values of the first two generation sfermion masses 
$m_0(2)$ vs. $m_0(1)$ (as $m_0(1,2)$ increase, then weak scale sfermion masses
are nearly equal to high scale sfermion masses). 
The line of degeneracy is solid black, while the bounds from
Misiak {\it et al.} are labeled in green. Here, we see that for
sparticle masses of order the weak scale, then rather strict degeneracy is required. However, as $m_0(1,2)$ increase, then degeneracy is gradually relaxed until
by $m_0(1,2)\sim 10$ TeV the bounds essentially disappear, showing again the
decoupling solution. In each of the four frames, we also show the 
predicted landscape distribution of sfermion masses for a statistical draw
of {\it a}) $n=1$, {\it b}) $n=2$, {\it c}) $n=3$ and {\it d}) $n=4$. 
We adopt particular, flavor-independent upper bounds of $m_0(1,2)<20$ and
40 TeV since the true upper bound is parameter dependent. 
In frame {\it a}) with $n=1$, just a few landscape points lie 
in the excluded region. As $n$ increases, then there is a stronger 
statistical draw towards large soft terms and the sfermion masses are drawn to
flavor independent upper bounds. Thus, there is also increasing degeneracy
of diagonal soft breaking terms. In this sense, the landscape provides a
mixed decoupling, quasi-degeneracy solution to the SUSY flavor problem.
For higher $n$ values, then none of the landscape points lie in the excluded region.

\subsection{CP}
\label{ssec:cp}

Limits can also be placed on complex valued soft terms due to their
inducement of CP violating effects on $\epsilon$ and $\epsilon^\prime /\epsilon$
in the kaon system and also from neutron ($d_n$) and electron ($d_e$)
electric dipole moments (EDMs)\cite{Dugan:1984qf,Gabbiani:1996hi}. 
The latter contribute only to LR mixing terms
and are suppressed by Yukawa couplings for the first two generations so we 
concentrate on the former kaon constraints.

\begin{figure}[!htbp]
\begin{center}
\includegraphics[height=0.42\textheight]{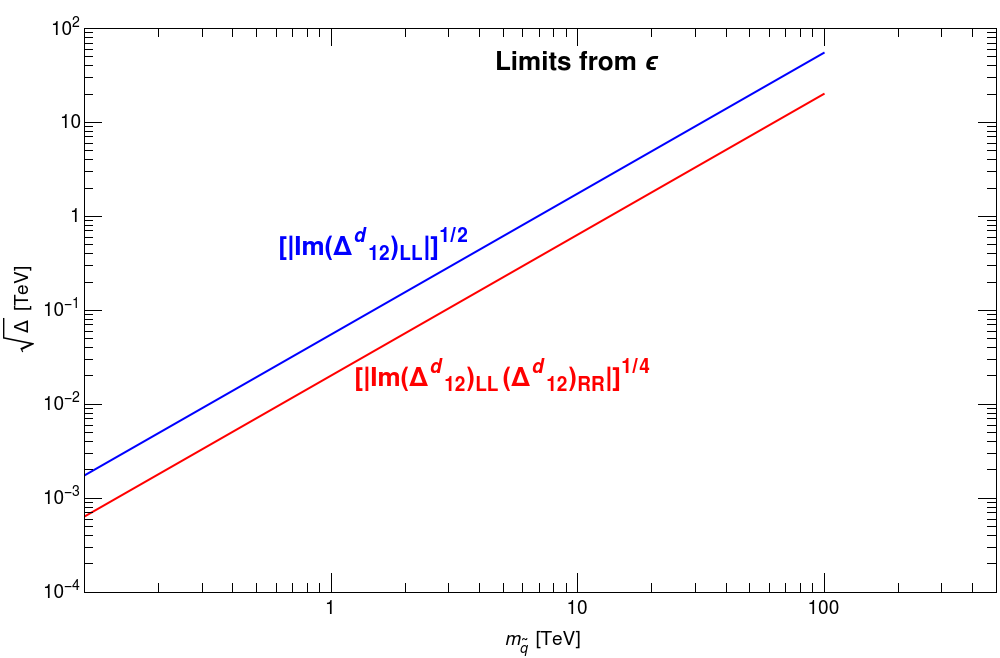}
\caption{Upper limits on $\left[Im|(\Delta_{12}^d)_{LL}|\right]^{1/2}$ (blue) 
and $\left[Im|(\Delta_{12}^d)_{LL}(\Delta_{12}^d)_{RR}|\right]^{1/4}$ (red)
from kaon system $\epsilon$ constraints.
\label{fig:eps}}
\end{center}
\end{figure}

In Fig. \ref{fig:eps}, we show the constraints on the Imaginary part
$[|Im(\Delta_{12}^d)_{LL}|]^{1/2}$ and \\
$[|Im(\Delta_{12}^d)_{LL}(\Delta_{12}^d)_{RR}|]^{1/4}$ from requiring contributions to the $\epsilon$ parameter to be below its 
measured value. The contributions are plotted against average
first/second generation squark mass for $m_{\tg}^2/m_{\tq}^2=0.3$. 
From the plot, we see that for weak scale sparticle masses 
$m_{\tq}\sim 100$ GeV, then the CP violating mass terms are required to be below about $0.5-2$ GeV. However, as $m_{\tq}$ is pulled towards the landscape
expected values in the tens of TeV range, then the CP-violating masses are
only constrained to be $\alt 4-10$ TeV (assuming 30 TeV squark masses).
For unfettered CP-violating soft masses, then squark masses are required
as high as 100 TeV.

\begin{figure}[!htbp]
\begin{center}
\includegraphics[height=0.42\textheight]{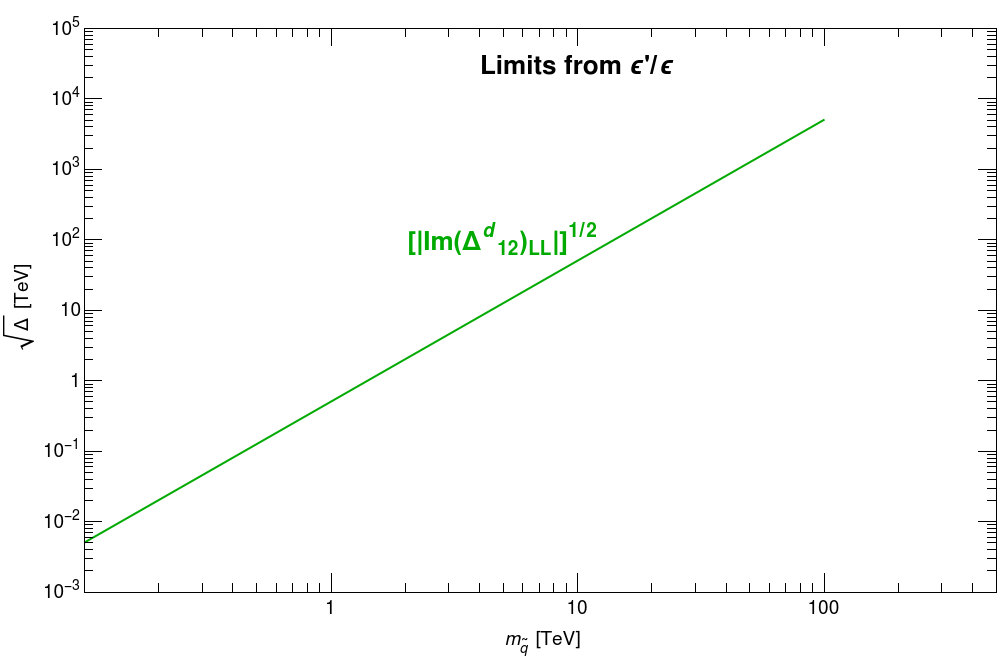}
\caption{Upper limits on Imaginary part of off-diagonal squark mass terms from
Kaon system $\epsilon^\prime /\epsilon$ constraints.
\label{fig:epsprime}}
\end{center}
\end{figure}

From the measured value of $\epsilon^\prime /\epsilon$, we can also
constrain $[ |Im (\Delta_{12}^d)_{LL}|]^{1/2}$. These results are shown
in Fig. \ref{fig:epsprime} versus the average first/second generation squark mass for $m_{\tg}^2/m_{\tq}^2=0.3$. 
For weak scale squark masses, then the CP-violating mass term is required
to be $\alt 5$ GeV. As $m_{\tq}$ increases into the expected landscape range
of $20-40$ TeV, then the CP-violating masses can lie in the 100 TeV range, 
thus solving the SUSY CP constraint at least in this channel.

\section{Conclusions}
\label{sec:conclude}

The string theory landscape provides a compelling picture for the magnitudes
of soft SUSY breaking terms. Rather general considerations of the string theory landscape from Douglas 
point to a statistical draw towards large soft terms while nuclear physics
calculations from Agrawal {\it et al.} require values of pocket-universe
weak scale value displaced by no more than a factor 2-5 from our measured 
value in order to produce atoms as we know them. 
Assuming a natural solution of the SUSY $\mu$ problem, 
with $\mu\sim m_{weak}$, then a 
statistical sampling of soft terms allows the calculation of pocket-universe
$m_Z^{PU}$ via Eq. \ref{eq:mzs}. 
The results of the statistical calculation pull $m_h\to 125\pm 2$ GeV 
while sparticle masses are pulled beyond present LHC reach. 
Only higgsinos need to lie close to the weak scale.

In this paper, we focused on the landscape pull on first/second generation 
sfermion masses. Their upper bound doesn't arise from EW $D$-term contributions
(which allow sfermions up to 1000 TeV due to large, nearly perfect 
cancellations). Instead, their upper bound arises from two-loop RG
contributions to third generation soft masses which actually push
these values to small, even tachyonic values. 
As shown in Fig. \ref{fig:m012},
this is yet another example of the landscape pull toward 
{\it living dangerously}: increasing first/second generation soft masses
make the theory increasingly natural until they move it towards disallowed 
too large weak scale values and ultimately to CCB minima in the 
Higgs potential. First/second generation soft masses are thus 
pulled into the tens of TeV range towards a flavor-independent upper bound.
This provides a mixed decoupling/quasi-degeneracy solution to the 
SUSY flavor and CP problems.

We evaluated FCNC and CP-violating constraints in Sec. \ref{sec:flavorCP}.
While the SUSY flavor and CP problems do require flavor universality for
weak scale sparticle masses, for sfermions in the tens of TeV range, then
the constraints are greatly weakened but not entirely destroyed. 
Typically, off diagonal soft term contributions to sfermion mass matrices
in the superCKM basis are required to lie in the multi-TeV region for
tens of TeV soft terms. In addition, the pull to large, quasi-degenerate
diagonal soft terms fulfills constraints on soft term degeneracy for 
the first two generations. Also, imaginary parts of SUSY soft terms
are only mildly constrained for sfermions in the 20-40 TeV range. 
As an example, we display in Table \ref{tab:sum} a summary of 
important constraints gained for the case of average sfermion mass
$m_{\tf}=30$ TeV.
Overall, we would conclude that the string landscape picture offers 
a compelling picture of at best only mild constraints on off-diagonal
flavor changing soft terms and CP-violating masses via a mixed
decoupling/quasi-degeneracy solution to the SUSY flavor problem 
and a decoupling solution to the SUSY CP problem.
\begin{table}[!htb]
\renewcommand{\arraystretch}{1.2}
\begin{center}
\begin{tabular}{c|c|c}
quantity & upper bound & source \\
\hline
$[|(\Delta_{12}^d)_{LL}(\Delta_{12}^d)_{RR}|]^{1/4}$ & $<12\ {\rm TeV}$ & $\Delta m_K$ \\
$[|(\Delta_{12}^d)_{LL}|]^{1/2}$ & $<30\ {\rm TeV}$ & $\Delta m_K$\\
$[|(\Delta_{12}^\ell)_{LL}|]^{1/2}$ & $<200\ {\rm TeV}$ & $BF(\mu\to e\gamma )$ \\
$|m_{\tq 1}-m_{\tq 2}|$ & unbounded & $\Delta m_K$ \\
$[|Im(\Delta_{12}^d)_{LL}|]^{1/2}$ & $<10\ {\rm TeV}$ & $\epsilon$ \\
$[|Im(\Delta_{12}^d)_{LL}(\Delta_{12}^d)_{RR}|]^{1/4}$ & $<3\ {\rm TeV}$ & $\epsilon $ \\
$[|Im(\Delta_{12}^d)_{LL}|]^{1/2}$ & $<500\ {\rm TeV}$ & $\epsilon^\prime /\epsilon$\\
\hline
\end{tabular}
\caption{Upper bounds from various measurements of flavor changing and 
CP violating quantities considered in text for 
average sfermion mass $m_{\tf}=30$ TeV.
}
\label{tab:sum}
\end{center}
\end{table} 

{\it Acknowledgements:} 
This work was supported in part by the US Department of Energy, Office
of High Energy Physics. 


%
\end{document}